\documentclass{emulateapj}

\usepackage{amsmath}
\usepackage{amssymb}
\usepackage{natbib}
\usepackage{xspace}
\usepackage{graphicx}
\usepackage{rotate} 
\usepackage{subfigure}
\usepackage{float}
\usepackage{rotate}

\newcommand{\err}[2]{\ensuremath{^{+#1}_{-#2}}\xspace}

\newcommand{\Msun}{\ensuremath{M_\odot}\xspace}

\newcommand{\NH}{$N_{\text{H}}$}
\newcommand{\NHgal}{$N_{\text{H\,Gal}}$}
\newcommand{\NHtor}{$N_{\text{H\,Tor}}$}
\newcommand{\mcg}{MCG\,--2-58-22}
\newcommand{\pexrav}{\textsc{pexrav}\xspace}
\newcommand{\mytorus}{MYT\textsc{orus}\xspace}
\newcommand{\etal}{et al.\,}
\newcommand{\Ka}{K$\alpha$}
\newcommand{\Kb}{K$\beta$}
\newcommand{\chidof}{$\chi^{2}/$dof\xspace}

\newcommand{\sax}{\textsl{BeppoSAX}\xspace}
\newcommand{\suzaku}{\textsl{Suzaku}\xspace}

\newcommand{\xmm}{\textsl{XMM-Newton}\xspace}

\shorttitle{\suzaku observation of \mcg}
\shortauthors{Rivers et al.}

\begin{document}

\title{A \textsl{Suzaku} Observation of \mcg:  \\ 
Constraining the Geometry of the Circumnuclear Material} 

\author{Elizabeth~Rivers\altaffilmark{1}, Alex~Markowitz\altaffilmark{1}, Richard~Rothschild\altaffilmark{1}}
\altaffiltext{1}{University of California, San Diego, Center for
  Astrophysics and Space Sciences, 9500 Gilman Dr., La Jolla, CA
  92093-0424, USA} 
\email{erivers@ucsd.edu}

\begin{abstract}

We have analyzed a \suzaku long-look of the active galactic nucleus \mcg, a type 1.5 Seyfert with very little X-ray absorption 
in the line of sight and prominent features arising from reflection off circumnuclear material: the Fe line and Compton reflection hump.  
We place tight constraints on the power law photon index ($\Gamma$=1.80$\pm$0.02), the Compton reflection strength ($R$=0.69$\pm$0.05), 
and the Fe K emission line energy centroid and width ($E$=6.40$\pm$0.02 keV, $v_\textrm{FWHM} <$\,7100 km\,s$^{-1}$).  We find no significant 
evidence for emission from strongly ionized Fe, nor for a strong, relativistically broadened Fe line, indicating that perhaps there is no 
radiatively efficient accretion disk very close in to the central black hole.  In addition we test a new self-consistent physical model
from Murphy \& Yaqoob, the ``\mytorus'' model, consisting of a donut-shaped torus of material surrounding the central illuminating source 
and producing both the Compton hump and the Fe K line emission.  From the application of this model we find that the observed spectrum is 
consistent with a Compton-thick torus of material (column density \NH=3.6\err{1.3}{0.8}$\times 10^{24}$ cm$^{-2}$) lying outside of the line 
of sight to the nucleus, leaving it bare of X-ray absorption in excess of the Galactic column.  We calculate that this material is sufficient 
to produce all of the Fe line flux without the need for any flux contribution from additional Compton-thin circumnuclear material.

\end{abstract}

\keywords{Galaxies: active -- X-rays: galaxies -- Galaxies: Individual: \mcg}

\section{Introduction}

\mcg\ is an X-ray bright Seyfert 1.5 active galactic nucleus (AGN) located at a redshift of $z$ = 0.04686.  
Past X-ray observations of this source performed with \textsl{EXOSAT}, \textsl{ASCA}, \xmm and \sax have revealed
the following spectral components in addition to the primary X-ray power law: a soft excess, Fe emission lines, and a Compton reflection hump.
Importantly, there has been no evidence for X-ray absorption by gas along the line of sight in
excess of the Galactic column, indicating that in the X-ray band this AGN is a ``bare nucleus''.  
This combination makes \mcg\ an interesting target of study, since the lack of significant X-ray absorption 
provides a clean view of the nucleus with a relatively simple spectum to model, while the presence of strong reflection components 
allows us to place constraints on the physical geometry of the circumnuclear material surrounding the AGN. 

Ghosh \& Soundararajaperumal (1992) analyzed \textsl{EXOSAT} data obtained in 1984 that revealed the highly variable soft 
excess below about 2 keV.  They modeled this component with a steep power law in addition to their continuum power law. 
\textsl{ASCA} data covering 2.5--10 keV with good CCD resolution were analyzed by Weaver \etal (1995).  
They modeled the spectrum using a hard X-ray power law ($\Gamma = 1.75 \pm 0.05$) with Galactic absorption and confirmed 
the need for a soft excess, as well as an Fe \Ka\ emission line which was unresolved. 
Weaver, Gelbord \& Yaqoob (2001) analyzed two additional \textsl{ASCA} observations of \mcg, tracking the Fe line flux
over a time scale of years and showing large variation in the flux of the underlying continuum. 
However large uncertainties precluded definitive conclusions about the variation of the Fe line parameters.

A more recent analysis by Bianchi \etal (2004) using simultaneous data from \xmm and \sax 
covered a much broader energy range than previous observations (0.5--200 keV).  Unfortunately only 7 ks 
of good EPIC-pn data were obtained for the source, providing only loose constraints in the Fe K bandpass 
(the Fe \Ka\ line was unresolved with $\sigma < 340$ eV and an equivalent width, EW, of 45\err{85}{24} eV).
They were able to loosely constrain the Compton reflection hump with 
$R = 0.4 \pm 0.3$ and $\Gamma = 1.72$\err{0.08}{0.06}.

In this paper we present an in-depth analysis of a single 140 ks long-look \suzaku observation of \mcg.  
\suzaku is a Japanese observatory that provides broadband X-ray spectra from $\sim$\,0.5 keV to above 500 keV 
with good energy resolution and effective area around 5--7 keV for detailed analysis of the Fe K complex (Mitsuda \etal 2007).
Our goals for this observation were to study the Fe K emission, constrain the Compton hump, confirm the 
bare nucleus, and study the soft excess, all of which \suzaku is capable of doing well.
This information can then be used to explore possibilities for the geometry of the 
circumnuclear material including the Fe line emitting gas and we improve upon parameter values in the literature 
for the Compton hump and Fe K emission lines.  
This paper is structured as follows: Section 2 details data reduction, Section 3 describes 
spectral fitting and analysis, and Section 4 contains a discussion of the results.


\begin{figure}
\plotone{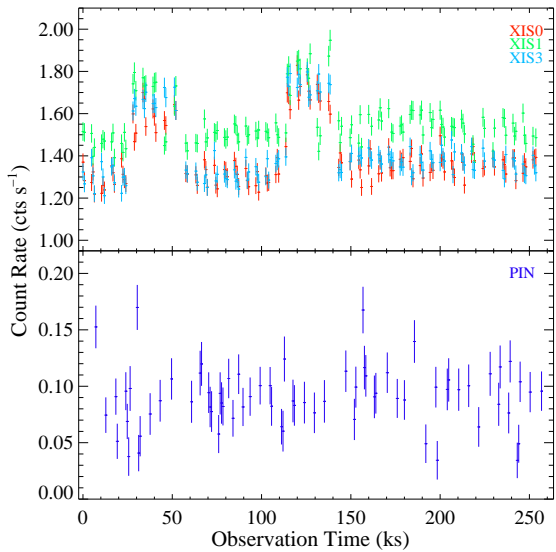}
\caption{The lightcurve over the duration of the observation.  Bins are 800\,s.  The XIS count rates are in the 2--10 keV 
range while the PIN count rates are in the 20--50 keV range.  The increases in flux seen in the XIS around 30--50 ks and 
110--140 ks are small (only about 20\%) and flux-resolved spectroscopy did not reveal significant change in the 
shape of the spectrum during these episodes.}
   \label{figlc}
\end{figure}

\section{Data Reduction and Analysis}\label{sec:analysis}

\textsl{Suzaku} observed \mcg\ with the X-ray Imaging Spectrometer (XIS; Koyama \etal 2007) and the 
Hard X-ray Detector (HXD; Takahashi \etal 2007) on 2009 November 27 beginning at 22:49 UT
(Observation ID 704032010).  Data were processed with version 2.4.12.27 of the \textsl{Suzaku} pipeline and 
typical screening criteria were applied (as per the \textsl{Suzaku} Data Reduction 
Guide\footnote{http://heasarc.gsfc.nasa.gov/docs/suzaku/analysis/abc/abc.html}).  
All extractions were done using HEASOFT v.6.9.


\subsection{XIS Reduction}

The XIS is comprised of 3 CCD cameras\footnote{The fourth CCD camera, XIS2, is inoperative as of 2006 November.  
See the \textsl{Suzaku} ABC Guide for details.} each placed in the focal plan of an X-ray Telescope module.
Two of these (XIS0 and XIS3) are front-illuminated (FI), maximizing the effective area of
the detectors in the Fe K bandpass, while the third (XIS1) is back-illuminated (BI), increasing its effective area 
in the soft X-ray band ($\lesssim$\,2 keV).  Two corners of each XIS CCD are illuminated by an $^{55}$Fe 
calibration source, which can be used to calibrate the gain and test the spectral resolution 
of data taken using this instrument (see the \textsl{Suzaku} Data Reduction Guide for details).

After screening, the good exposure time per XIS was 138.9~ks.
The XIS events data were in 3$\times$3 and 5$\times$5 editing modes which were cleaned and summed 
to create image files for each XIS.   
From these we extracted source and background lightcurves and spectra, using XISRMFGEN and XISSIMARFGEN 
to create the response matrix (RMF) and ancillary response (ARF) files. 
Data from the two front-illuminated CCDs were summed to create a single co-added FI spectrum after it was
confirmed that the two spectra were consistent.

Data were ignored above 12 keV (10 keV for BI) where the effective area of the XIS begins to drop dramatically. 
Data were ignored below 1.0 keV (0.7 for BI which has a larger effective area at low energies) 
due to time-dependent calibration issues of the instumental O K edge at 0.5 keV,
and between 1.5 and 2.4 keV due to large calibration uncertainties for the Si K complex and 
Au M edge arising from the detector mirror system. These issues are not fully understood at the time of this writing.
Average 2--10 keV rates were 1.410$\pm 0.002$ and 1.521$\pm 0.003$ counts\,s$^{-1}$ per XIS for FI and BI respectively.
Figure \ref{figlc} shows the XIS lightcurves for the duration of the observation. 

Fitting the $^{55}$Fe calibration source spectra in XSPEC v.12.6.0 (Arnaud \etal 1996) with a model comprised of 
three Gaussian components (Mn K$\alpha_{1}$, K$\alpha_{2}$ and K$\beta$) yielded the following 
results for the Mn K$\alpha_{1}$ line energy (expected value of 5.899 keV):  5.886 keV (FI) and 5.890 (BI),
showing that the energy calibration has a systematic untertainty of $\sim$\,10 eV for both the FI the BI.  
Additionally, these lines had an average width of 30 eV, which we will take as instrumental broadening in 
excess of that modeled by the response matrix, and have subtracted this value in quadrature from all measured
line widths.

\subsection{PIN Reduction}

The HXD gathered data with both its detectors, the PIN diodes 
and the GSO scintillators, however we did not use the GSO data because of the faintness of 
the source relative to the non-X-ray background in the GSO band.  
The HXD/PIN is a non-imaging instrument with a 34$\arcmin$ square (FWHM) field of view.  
The HXD instrument team provides non-X-ray background model event files using the calibrated GSO 
data for the particle background monitor (``background D'' or ``tuned background'' with METHOD=LCFITDT).  
This yields instrument background estimates with $\lesssim$\,1.5\% systematic uncertainty at the 1$\sigma$\ 
level (Fukuzawa \etal 2009).  As suggested in the Suzaku ABC Guide, the Cosmic X-ray Background 
was simulated in XSPEC v.12.6 using the form of Boldt (1987).

Net spectra were extracted and deadtime-corrected for a net exposure time of 98.0 ks.  
We excluded PIN data below 13 keV due to thermal noise and above 60 keV where the effective area
of the detector falls significantly.  
The average 13--60 keV rate was 0.202$\pm 0.002$ counts\,s$^{-1}$.
Figure \ref{figlc} shows the  PIN lightcurve for the duration of the observation. 


\section{Spectral Fitting}

All spectral fitting was done in XSPEC, utilizing solar abundances of
Anders \& Grevesse (1989) and cross-sections from Verner \etal (1996). 
All fits included absorption by the Galactic column with \NHgal=2.70$\times 10^{20}$\,cm$^{-2}$ (Kalberla \etal 2005).
Uncertainties are listed at the 90\% confidence level ($\Delta \chi^2$ = 2.71 
for one interesting parameter).


\begin{figure}
  \plotone{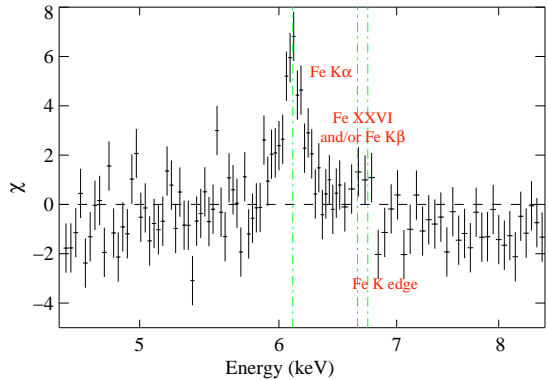}
  \caption{Data--model ratios to a simple power law fit for the Fe K band from 4.5--8.5 keV using the FI XIS data only.
Dashed lines show expected locations (\textsl{from left to right}) of the energy centroids of the Fe \Ka, XXVI, and \Kb lines. 
Note that these are the observed energies ($z$ = 0.04686).}
   \label{figfek}
\end{figure}

\begin{deluxetable*}{lccccccccccc}
   \tablecaption{Model Parameters \label{tabbase}}
   \tablecolumns{12}
\startdata
\hline
\hline\\
Model    &     $\Gamma$   &  $A$\tablenotemark{a} & $E$  &   $I$\tablenotemark{b}  & $\sigma$ &  EW &  $\tau$  &  $R$  &  $\Gamma_\textrm{soft}$ & $A_\textrm{soft}$\tablenotemark{a} & $\chi^{2}$/dof\\[1mm]
&       &      ($10^{-2}$)       & (keV)     & ($10^{-5}$) &  (eV) & (eV)& & &&($10^{-3}$)&\\[1mm]
\hline\\[0.3mm]

Fe K Band   &  1.68$\pm 0.05$   &   1.15 $\pm 0.08$  & 6.40$ \pm 0.02$  & 2.8 $\pm 0.6$ & 60 $\pm 30$ & 50$\pm 10$ & 0.05$\pm0.02$  &  &  &  &  103/95\\[1mm]


\textbf{Broadband 1}  &  1.80$\pm 0.02$   &   1.30$\pm 0.02$  & 6.40$ \pm 0.02$  & 2.4 $\pm 0.3$ & $< 65$ & 40$\pm 10$ & & 0.69$\pm 0.05$  &  3.0$\pm 0.6$   &  0.14\err{0.16}{0.07}  &  604/464   \\[2mm] 
\hline\\[0.3mm]
&&&&&&&&& $k_\textrm{B}T$ (keV)  & $A_\textrm{bbody}$\tablenotemark{c} &\\[1mm]
Broadband 2  &  1.83$\pm 0.01$   &   1.37$\pm 0.01$  & 6.40$ \pm 0.02$  & 2.4 $\pm 0.3$ & $< 60 $ & 41$\pm 5$ & & 0.76$\pm 0.07$  & 0.18 $\pm 0.02$   & 2.2 $\pm 0.5$  &  611/464\\[-1mm]

\enddata

\tablecomments{Best fit parameters for models in the Fe K bandpass and broad band spectrum.  ``Fe K Band 1'' is a model fit over the energy range 4.5--8 keV 
using only XIS1 (as described in Section 3.1) including the primary power law, Fe \Ka\ and \Kb\ lines as Gaussians and the Fe K edge 
(``\textsc{zedge}'' in XSPEC).  The Broadband models were fit over the energy range 0.7--50 keV using FI and BI XIS data as well as the PIN.
Broadband 1 includes the continuum, Gaussian Fe emission lines, the Compton hump modeled with \pexrav, and the soft excess modeled with a power law
as described in Section 3.2.  Broadband 2 differs from Broadband 1 in the use of a blackbody (``\textsc{bbody}'') to model the soft excess.
We adopt the parameters of the best fit to the Broadband 1 model for the discussion in Section 4.}

\tablenotetext{a}{Power law normalization (ph\,keV$^{-1}$\,cm$^{-2}$\,s$^{-1}$ at 1 keV)}
\tablenotetext{b}{Fe \Ka\ line normalization (ph\,cm$^{-2}$s$^{-1}$)}
\tablenotetext{c}{Blackbody normalization (10$^{-5}$ ph\,keV$^{-1}$\,cm$^{-2}$\,s$^{-1}$)}
\end{deluxetable*}


\subsection{The Fe K Bandpass}

We began our analysis with a preliminary study focused on the Fe K bandpass. 
We used data from 4.5--8.5 keV from the FI spectrum only because of its excellent response and effective area in this energy range.
We analyzed the Fe K complex, including the Fe \Ka\ and \Kb\ lines and the Fe K edge, and 
investigated the possiblity of emission from ionized Fe, namely Fe XXV or XXVI (the latter was reported by
Bianchi \etal 2004 with a 2$\sigma$ detection).  

As a first step we fit a simple power law with Galactic absorption.  Model--data residuals for the simple power law 
are shown in Figure \ref{figfek}.  This yielded a poor fit with \chidof = 322/100 and obvious residuals around 6.4 keV 
(rest frame energy), the location of the Fe \Ka\ line.  Fitting the line with a Gaussian component provided a much better 
fit with \chidof = 117/97.   Visual inspection then revealed additional residuals around 7.1 keV (the location of the \Kb\ line 
and Fe K edge). 

We then added an edge component to model additional Fe K shell absorption in excess of the Galactic
column and/or the Fe K edge associated with Compton reflection.  We fixed the edge energy 
at 7.11 keV and left the optical depth ($\tau$) free.  \chidof dropped to 108/96, indicating a significant
detection of the edge at a confidence level of $\sim$\,99.4\% according to an $F$-test 
\footnote{Note that an $F$-test is inappropriate to perform in this case (see Protassov \etal 2002), however it can 
give a rough approximation of the significance.}.

Assuming an origin in neutral or lowly-ionized gas, an Fe \Kb\ line should be present in addition to the Fe \Ka\ line.
We added a Gaussian emission line with its energy centroid frozen at 7.056 keV (degeneracy with the Fe K edge 
at 7.11 keV and lack of sufficient line strength to provide good constraints led to our freezing the parameter at 
its expected value), its width tied to that of the \Ka\ line and its normalization left free.
The fit improved, with \chidof dropping to 103/95 with a normalization of 15$\pm$12\% of the \Ka\ normalization,
consistent with that expected for cold/neutral gas.  An $F$-test$^3$ indicates that this is a 2$\sigma$ detection
at the $\sim$\,96.4\% confidence level.

In some AGN, contributions to the total observed Fe K emission profile can arise from material which is ionized, either
by collisional- or photo-ionization.  Using \xmm-EPIC data, Bianchi \etal (2004) found a degeneracy between 
the parameters of the \Kb\ line and those from a possible Fe XXVI emission line.  
They reported a 2$\sigma$ detection of the Fe XXVI line when the \Kb\ line was not included
and with all the parameters of the line left free.  When we allowed the energy of the \Kb\ line to be free to vary we found an
energy centroid of 7.0$\pm 0.1$, consistent with both the \Kb\ and Fe XXVI line energies.  \chidof was 102/95 (not a significant 
improvement) and the normalization was 18$\pm$14\% of the \Ka\ normalization.  Freezing the line energy at 6.966 keV, the
weighted average of the Fe XXVI doublet, provided a fit virtually identical to the one presented above.
Fitting both lines simultaneously with energies frozen at their expected values and widths tied to that of the \Ka\ line gave 
\chidof=102/94 with a normalization of 13\err{18}{11}\% for Fe XXVI and an upper limit of 26\% for \Kb\ (both percentages are with
respect to the \Ka\ normalization).
We therefore cannot rule out the that the source may contain both emission lines and that we are simply unable to deblend them.
For simplicity, in all further models described in this paper we have included only the \Kb\ line with frozen or tied parameters.

We also tested for the presence of Fe XXV (using a Gaussian component with energy centroid fixed at 6.70 keV), however \chidof did 
not improve and only an upper limit to the normalization was obtained ($\lesssim$\,6$\times 10^{-6}$ ph\,cm$^{-2}$s$^{-1}$).
Final parameters for the Fe K complex model fit including the \Ka\ line, Fe K edge and \Kb\ line are listed in Table 1 as 
the ``Fe K Band'' model, including the Fe \Ka\ energy centroid ($E$), intensity ($I$), width ($\sigma$), and EW.  

We also tested the ``\textsc{diskline}'' model for the emission lines in place of the more phenomenological Gaussian model.
\textsc{diskline} models the Doppler broadening of an emission line associated with the inner region of an accretion disk (Fabian \etal 1989).  
The diskline parameters were not well constrained, giving a very large inner radius (R$_\textrm{in} \gtrsim$\,20 $R_{\rm S}$, 
where $R_{\rm S}$ is the Schwarzchild radius) and a narrow profile.  It did not improve the fit over a simple Gaussian.
Next we tried the addition of a broad \textsc{diskline} (R$_\textrm{in}$ constrained to around 3 $R_{\rm S}$ and energy fixed at 6.4 keV) 
to a narrow Gaussian line with  energy and width fixed at the values found with the Gaussian fit (see Table 1).  We obtained an upper limit 
to the \textsc{diskline} normalization of $\lesssim$\,2$\times 10^{-5}$ ph\,cm$^{-2}$s$^{-1}$ and an EW of $\lesssim$\,35 eV.  
It is thus possible that a weak broad line exists in this source and that we are simply unable to detect it; however combining the weakness
of this feature with the lack of ionized emission indicates that most if not all of the Fe line flux comes from material that is 
not close in to the central black hole.

It is possible that such a weak broad line would be degenerate with the Compton shoulder, a feature which could arise if there
is a significant amount of Compton-thick material.
Additionally, by visual inspection we see a shallow shelf-like shape in the residuals on the low-energy side of the 
emission line when $\sigma$ is set to $\lesssim$\,30 eV.
We tested this by modeling a moderately broad Gaussian in addition to the narrow Fe \Ka\ line
in the ``Fe K Band 2'' model with an energy centroid fixed at 6.34 keV (Matt 2002).  This resulted in an improvement in \chidof 
of only 2/1, which is not a significant detection but indicates that the Compton shoulder should be tested for in 
future observations.


\begin{figure}
  \plotone{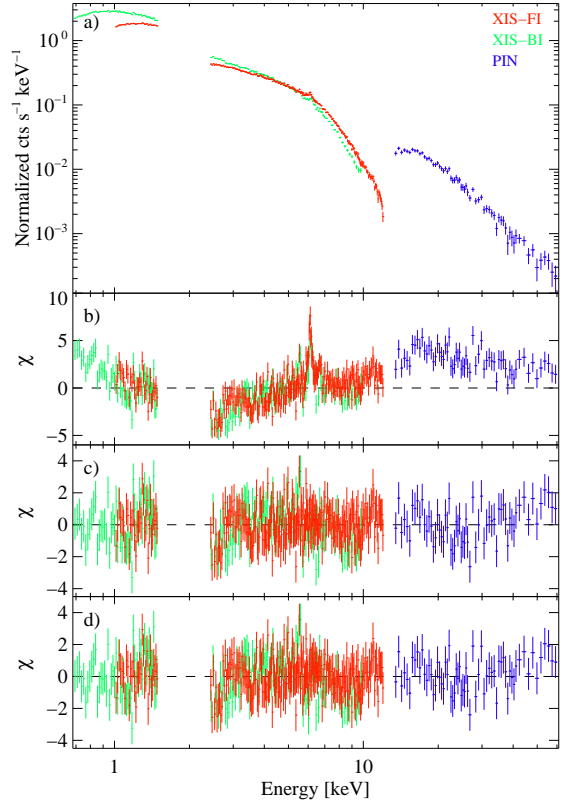}
  \caption{Spectral fitting for \mcg\ from 0.7--50 keV.  Panel a) shows the data from the XIS FI and BI, and the PIN.  
Panel b) shows the data--model residuals for a simple absorbed power law.  Panel c) shows the data--model residuals 
for our best-fit disk/slab geometry model including the iron lines, soft excess and Compton hump in addition to the continuum.
Panel d) shows the data--model residuals for our best fit torus geometry model utilizing \mytorus.  It is obvious that the fits
shown in panels c) and d) are virtually identical though the modelling is quite different.}
   \label{figspec}
\end{figure}

\begin{figure}
  \plotone{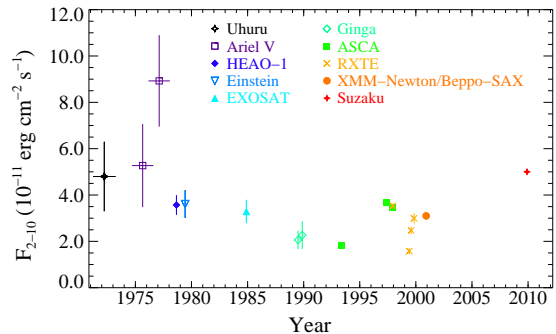}
  \caption{Historical values for the 2--10 keV flux.  From left to right this source has been observed by \textsl{Uhuru} 
(Cooke \etal 1978), \textsl{Ariel-V} (Marshall, Warwick \& Pounds 1981), \textsl{HEAO-1} (Griffiths \etal 1979), \textsl{Einstein} 
(Turner \etal 1991), \textsl{EXOSAT} (Ghosh \& Soundararajaperumal 1992), \textsl{Ginga} (Nandra \& Pounds 1994), \textsl{ASCA} 
(Weaver \etal 1995), \textsl{RXTE}, simultaneously by \xmm/\sax (Bianchi \etal 2004), and most recently by \suzaku (this paper).
}
   \label{figltlc}
\end{figure}


\subsection{Broadband Fitting}

Next we fit the broadband spectrum of \mcg\ covering the range from 0.7--60 keV.  
We used the XIS FI, XIS BI, and PIN spectra.  We included an instrumental cross-normalization constant in our fits with
the PIN constant set to 1.16 (this is the expected value for XIS-nominal pointing and leaving the parameter 
free caused degeneracy with the Compton reflection component in our models) and the BI constant left free relative 
to the FI spectrum (values were typically around $\sim$1.05).  The broadband data are shown in Figure \ref{figspec}a.
Figure \ref{figspec}b shows residuals to a simple power law fit in which we can clearly see the need for modeling the 
Fe K complex, Compton reflection peaking around 20--30 keV, and a soft excess below $\sim$\,1 keV.

We began by modeling the soft excess with a simple power law in addition to the continuum power law and modeling the Compton
reflection hump using \pexrav (Magdziarz \& Zdziarski, 1995) which assumes a disk-like geometry for the reflecting material and
where the value of $R$ is the proportion of the primary power law that is reflected off Compton-thick material.  
Best fit parameters for this model are listed under ``Broadband 1'' in Table 1 and data--model residuals are shown in
Figure \ref{figspec}c.  We did not find the need for additional 
cold absorption with an upper limit to the column density of 2.5$\times 10^{20}$ cm$^{-2}$ in excess of the Galactic.
We also tested for the presence of warm absorption using an \textsc{xstar} table but found it unnecessary for a good fit.
The Compton reflection hump and Fe K edge were well fit by the \pexrav model and Fe line parameters were very similar to 
those found from narrow band fitting, including the upper limit to a relativistically broadened \textsc{diskline} component.
The observed 2--10 keV flux was F$_{2-10}$=5.0 $\pm 0.1 \times 10^{-11}$ ergs\,cm$^{-2}$\,s$^{-1}$
and the intrinsic luminosity, calculated using the cosmology-corrected luminosity distance given by the NASA/IPAC Extragalactic 
Database\footnote{http://nedwww.ipac.caltech.edu/}, and correcting for Galactic absorption was L$_{2-10}$=2.30$\pm 0.05 \times 10^{44}$ ergs\,s$^{-1}$.
Figure \ref{figltlc} shows historical values of the flux of this source in the 2--10 keV range.

We also tried a (phenomenological) blackbody emission component to model the soft excess.  Parameters are listed in Table 1
under ``Broadband 2''.  Both the blackbody and power law models fit the data reasonably well, but in both cases parameters were 
difficult to constrain due to calibration issues with the O K edge below about 1 keV.
For simplicity we adopt Broadband 1 as our best-fit model for the discussion in Section 4.

Bianchi \etal (2004) tested for high energy cut offs in their sample finding model-dependent values of $E_{\rm c} \sim$ 200 keV with 
error bars of 50--800 keV.  Extending up to only 60 keV, our data are not highly sensitive to a high energy cut off or rollover. 
Utilizing the ``\textsc{cutoffpl}'' model in XSPEC we found a lower limit to the rollover energy of 500 keV.


\begin{deluxetable*}{lccccccc}
   \tablecaption{Model Parameters for \mytorus \label{tabmyt}}
   \tablecolumns{8}
\startdata
\hline
\hline\\
$\Gamma$   &  $A_\textrm{PL}$\tablenotemark{1} & \NHtor  &   $A_{\rm L}$  & $\sigma$ &  $\Gamma_\textrm{soft}$ & $A_\textrm{soft}$\tablenotemark{1} & $\chi^{2}$/dof\\[1mm]
     &      ($10^{-2}$)       &  ($10^{24}$\,cm$^{-2}$)    &  & (eV)  &  &  ($10^{-3}$)&\\[1mm]
\hline\\[0.3mm]

    1.70$\pm 0.01$   &   1.13$\pm 0.03$  & 3.6\err{1.3}{0.8}  & 0.75 $\pm 0.14$ &  $<$70 & 2.5$\pm0.3$ & 0.4\err{0.4}{0.2} &  623/465\\[-1mm]

\enddata

\tablecomments{Best fit parameters for the self-consistent \mytorus model with $\theta_{\rm incl}$ fixed at 30$\degr$
as discussed in Section 3.3.}

\tablenotetext{1}{Power law normalization (ph\,keV$^{-1}$\,cm$^{-2}$\,s$^{-1}$ at 1 keV)}

\end{deluxetable*}


\subsection{Applying a Self-Consistent Model}

As our knowledge of AGN improves, so too should the sophistication of our modeling.  Self-consistent models
should be able to simultaneously model absorption and reflection by circumnuclear material, combining the
Fe line, Compton reflection hump and column density along the line of sight.  We have applied
the model ``\mytorus'' (Murphy \& Yaqoob 2009) to our spectrum of \mcg\ which assumes the circumnuclear material
is a donut shape of uniform density and includes all three of the components listed above.

The \mytorus model was derived from Monte Carlo simulations of a dusty torus of uniform density 
surrounding an illuminating supermassive black hole.  Relevant parameters to the model include the following: \NHtor, the column 
density of the material in the torus (not necessarily in the line of sight); $\theta_{incl}$, the inclination angle of the torus,
with 0$\degr$ corresponding to a face-on view, 90$\degr$ corresponding to edge-on, and with the torus intersecting the line of
sight for angles larger than 60$\degr$ (the assumed half-opening angle); the photon index ($\Gamma$) and the
normalization ($A_\textrm{PL}$) of the illuminating power law; the width ($\sigma$) of the Fe \Ka\ line (the material is assumed to be
cold and the energy is not a free parameter); and with additional parameters $A_{\rm S}$ and $A_{\rm L}$, the normalization factors of 
the Compton hump and Fe line respectively, to be used when the amount observed for either is significantly different 
from that expected by the model due to differences in covering factors, abundances, etc., from those assumed.  
We also included an additional power law for the soft excess.

We found a reasonably good fit (\chidof = 623/465, very similar to the broadband fit $\chi^2$ values given in Table 1) 
with $\theta_{\rm incl}$ fixed at 30$\degr$ (there was no significant improvement in fit with this parameter free).
Since there is no extra absorption in \mcg, the value obtained for \NHtor ~is driven primarily by the strength of the 
Compton reflection hump and Fe line.
When $A_{\rm L}$ was free to vary there was an improvement in $\chi^2$ of 9 for 1 less degree of freedom.
We obtained a value for $A_{\rm L}$ 0f 0.75$\pm 0.14$, that is the amount of material creating the Fe line was about 
75\% of that creating the Compton hump, possibly due to an underabundance of Fe or geometrical effects not taken into account
by the model (it should also be noted that the upper uncertainty on $A_{\rm L}$ is consistent with the lower uncertainty on \NHtor).  
Testing for an additional relativistically broadened Fe line yielded an upper limit to the normalization of 
$\lesssim$\,1.5$\times 10^{-5}$ ph\,cm$^{-2}$s$^{-1}$ and an EW of $\lesssim$\,30 eV.
Our best fit parameters are listed in Table \ref{tabmyt} and data--model residuals are shown in Figure \ref{figspec}d.  


\section{Discussion and Conclusions}

\subsection{The Fe K Complex}
Focusing on the Fe K band we found the need for both Fe \Ka\ and \Kb\ emission lines as well as an Fe K
shell absorption edge (in broadband fits this edge was modeled sufficiently by the edge associated with the
Compton reflection hump in both \pexrav and \mytorus).
From the value of the emission line width found in our best-fit broadband model we calculated the 
velocity full width at half maximum ($v_\textrm{FWHM}$) of the 
emitting material to be $<$\,7100 km\,s$^{-1}$.  This is consistent with values obtained for
the optical H$\beta$ broad emission line of around 6400--8500 km\,s$^{-1}$ (Osterbrock 1977; 
Kollatschny \& Dietrich 2006; Winter \etal 2010) and is a significant improvement on previous upper limits set
by Weaver \etal (1991) and Bianchi \etal (2004) of $\lesssim$\,30,000 km\,s$^{-1}$.
Using a black hole mass estimated from optical luminosity and line widths to have a value of $10^{8.4}$ \Msun 
(Bian \& Zhao 2003; Winter \etal 2010) and assuming Keplerian motion of the 
emitting material, we estimated the radius of the emitting region to be\ $\gtrsim$\,45 lt-days 
or roughly 1200 $R_{\rm S}$.  

We also tested for a broad line and Compton shoulder.  According to de la Calle P\'erez \etal (2010),
roughly $\gtrsim$\,1.5$\times 10^5$ counts in the 2--10 keV band at CCD resolution provide good enough statistical quality to 
significantly detect a broad line.  In the combined FI XIS we have $\sim$\,4$\times 10^5$ counts, 
and our upper limit on the EW of a broad line places us
in the lower part of the EW range of detected broad lines
in the FERO sample of Seyferts observed with \textit{XMM-Newton}, wherein
significant detections of broad lines with EW's in the range of 50--250 eV were reported.
We conclude that a very strong broad line ($\gtrsim$\,50 eV) does not exist in MCG--2-58-22
or else our observation would have been sufficient to significantly detect it;
if there does exist a broad line in this source, then it must be very weak.

We would expect to see a Compton shoulder given the presence of Compton-thick material, however we did not obtain a 
significant detection.  It should also be noted that the Compton shoulder is included in the \mytorus model automatically, 
based on the strength of the Fe line and the column density of the torus.

\subsection{Reflection and Geometry of the Circumnuclear Material}
This source also shows a very prominent Compton hump around 20--30 keV (see Figure \ref{figspec}b).  This feature, arising from
Compton scattering of high energy photons off Compton thick material in the vicinity of the black hole,
is often associated with the Fe line emission seen in AGN, since the same Compton thick material that 
produces the Compton hump also produces Fe K emission.  
By knowing the (model-dependent) relationship between the strength of the Compton hump ($R$) and the expected
Fe line EW, we can test if Compton thick gas is capable of accounting for the entire observed Fe line flux.  

Based on calculations done by George and Fabian (1991) for a disk geometry, we found that the expected Fe line EW
for our value of $R$ (assuming an inclination of 30$\degr$ and solar abundances) is $\sim$\,80$\pm 6$ eV.  In our
broadband fits we see a considerably milder Fe line flux with an EW closer to 40 eV, about half of the expected value.
We found a similar, though less robust result using \mytorus, with $A_{\rm L} =$ 0.75$\pm 14$ and EW = 47$\pm 6$ eV (calculated from the flux of the line)
which assumes a ``donut-like'' geometry, in this case a Compton-thick torus out of the line of sight to the nucleus.
In both cases we have found a lower than expected Fe line EW, possibly due to an underabundance of Fe.
This makes it unlikely that there is significant contribution to the Fe line flux from Compton thin material in the vicinity of the black hole.
This is in contrast to many Seyferts; for example the sample of Rivers \etal (2011), which utilized the \pexrav 
disk reflection model, found that on average only $\sim$30\% of the Fe line flux in Seyferts is associated with the Compton reflection 
component, implying the presence of substantial amounts of Compton-thin material and/or supersolar abundances of Fe.
However other samples of Seyferts observed with \sax and \xmm (Perola \etal 2002; Bianchi \etal 2004) were shown to have
values of EW and $R$ consistent with the Fe line and reflection hump arising in the same material assuming a slab (disk) geometry 
and a torus geometry respectively.  In all cases uncertainties have been quite large and further investigation is warranted.

\subsection{Optical Obscuration and X-ray Absorption}

We have confirmed that \mcg\ is unabsorbed in the X-ray band, a fact which is interesting considering its 
optical classification as a Seyfert 1.5 (Winkler 1992; Winter \etal 2010).  Standard unification schemes would suggest that
it should have less obscuration in the X-ray band than a typical Seyfert 2 but \textsl{more} than a typical Seyfert 1.   
However this is not the case, as we have obtained a very low upper limit on column density in excess of 
the Galactic.  The question then, is whether there is material in the line of sight to the optical emission
from this AGN that is \textsl{not} in the line of sight to the X-ray emitting region.

Optical reddening from dust can be characterized by the visual extinction ($A_\textrm{v}$), which we calculated 
from the flux ratio of H$\alpha$ to H$\beta$, using the observed values of Winkler (1992) and assuming the intrinsic value of the Balmer 
decrement to be 2.87 (Osterbrock 1989).  We found $A_\textrm{v}$\,$\sim$\,1.97~mag. 
From this value we calculated the inferred column density of gas using the relation of Predehl \& Schmitt (1995),~
\NH\,=\,$(1.79 \times 10^{21}$ cm$^{-2})$\,$A_\textrm{v}$, and assuming the Galactic gas/dust ratio holds.  
We inferred a column density of $\sim$\,3.55$\times 10^{21}$ cm$^{-2}$, an order of magnitude higher than our upper 
limit on the observed X-ray absorption column (in excess of the Galactic) of 2.5$\times 10^{20}$ cm$^{-2}$.  
Therefore, assuming the X-ray absorbing gas and optical absorbing dust track each other, the dust obscuring the optical broad line 
region is most likely \textsl{not} in the line of sight to the X-ray emitting regions.  The inferred column density is also far too low to be 
associated with the dusty torus out of the line of sight which must be Compton-thick in order to produce the strong reflection component 
that we see (\mytorus gives a column density of 3.6\err{1.6}{0.9}$\times 10^{24}$ cm$^{-2}$).  

Since the X-ray emitting region is theorized to be very close in to the central black hole, it seems unlikely that this dust is in the form of 
an extended cloud in the host galaxy lying far (kiloparsecs) from the black hole that just happens to have a hole in the right place to produce a 
bare nucleus in the X-rays.
The dust must be distinct, however, from the Compton-thick torus, given the low inferred column density
and evidence that we are viewing the AGN more or less face on.
One possible explanation for this set of constraints is that there is clumpy material in the line of sight and it happens that none
of the clumps are obscuring the X-ray emitting region (see, e.g., Nenkova \etal 2008), only the optical broad line region.  
If future X-ray monitoring discovers a sudden, short-term increase in \NH, this scenario may be supported. Another explanation is that 
the material is commensurate with the broad line region, thus obscuring only this region.


\subsection{Conclusions}

\suzaku is an ideal tool for studying AGN, providing us with a detailed look at many crucial components of AGN spectra. 
We confirmed that \mcg\ is extremely unabsorbed in the X-ray band (an upper limit of 2.5$\times 10^{20}$ cm$^{-2}$ in excess 
of the Galactic column), despite significant reddening seen in the optical band.
These results led us to conclude that while no absorbers are in the line of sight to the central region of the AGN, it is
possible that clumps of material may be obscuring lines of sight to the region(s) where the optical line emission is produced.

With the excellent resolution of the XIS CCDs around $\sim$\,5--7 keV we were able to accurately study the Fe K complex. 
We found a narrow ($v_\textrm{FWHM} <$\,7100 km\,s$^{-1}$) Fe \Ka\ emission line and were able to 
constrain the location of the emitting material to much farther out than has been done with previous observations.  
We detected no significant broad Fe line as we would expect to see from the inner portions of a radiatively efficient accretion disk.
Since we have such a clean line of sight to the nucleus, this region cannot be simply obscured or out of our line of sight, 
implying that the inner disk may be truncated or radiatively inefficient.  From our limits on the Fe line $v_\textrm{FWHM}$ we have 
calculated a minimum inner radius of $\gtrsim$\,1200 $R_{\rm S}$.  We were also able to associate
the Fe emission complex with the Compton reflection component in our models.  Our results indicate that both components likely
arise in the same Compton-thick material without any contribution from additional Compton-thin circumnuclear material.
Assuming a disk/slab geometry (the \pexrav model in XSPEC) for the Compton-thick material gave a reflection strength $R$=0.69$\pm$0.05 with a 
photon index of $\Gamma$=1.80$\pm$0.02.

We successfully applied the new \mytorus model for Compton reflection, which assumes that the reflecting material is in the form of a torus of 
uniform density rather than in the form of a flat disk, as has typically been done with more established models such as \pexrav.
Our results gave a photon index of $\Gamma$=1.70$\pm$0.01 with a column density of \NH=3.6\err{1.3}{0.8}$\times 10^{24}$ cm$^{-2}$ for the torus. 
Additionally we found a lower than expected normalization for the Fe line, possibly due to subsolar abundances of Fe with Z$_{\rm Fe}$=0.75$\pm$0.14.

\begin{acknowledgments}
This research has made use of data obtained from the \textsl{Suzaku} satellite, a collaborative 
mission between the space agencies of Japan (JAXA) and the USA (NASA).
This work has made use of HEASARC online services, supported by NASA/GSFC, and the NASA/IPAC Extragalactic Database, 
operated by JPL/California Institute of Technology under contract with NASA.
This research was supported by NASA constract NAS5-30720, and NASA grants NNX08AD72G and NNX10AH87G.
\end{acknowledgments}



\end{document}